# Learning Effects in Concept Association: A Study with Multidimensional Scaling

Hyago Ribeiro Torres Felix, Geovani Ferreira Barbosa, Jusciane da Costa e Silva, Gustavo de Oliveira Gurgel Rebouças, and Carlos Alberto dos Santos[1]*

Federal Rural University of the Semi-Arid Region
Department of Natural Sciences, Statistics and Mathematics
Av. Francisco Mota, Mossoró, RN, Brazil

[1] *Corresponding author: cas.ufrgs@gmail.com*

**Abstract**
This work presents a study designed to investigate learning effects in the association of scientific concepts. In the experiment, high school students worked with advance organizers, as proposed by Ausubel, and with an Arduino board to play songs they had programmed. Numerical Association of Concepts Tests (TANC†), administered before and after the instructional intervention, were analyzed using ALSCAL, a multidimensional scaling program available in the SPSS statistical package. The results strongly suggest that TANC is sensitive to changes in students' cognitive structures and that ALSCAL is capable of extracting structural information from the tests.

## 1. Introduction

For more than a century, word-association tests have been used in studies of psychological behavior (MOORE; GILLILAND, 1921; SMITH, 1922) and cognition (BROWN; BERKO, 1960; DEESE, 1959; MCGEHEE, 1937). From the 1960s onward, these tests began to be used in physics education research (JOHNSON, 1964, 1967; SHAVELSON, 1973), with an innovative computational approach whose foundations had been presented in 1952 (TORGERSON, 1952). Torgerson's foundations were subsequently brought into the computational framework familiar today (KRUSKAL, 1964; SHEPARD, 1962a, b), giving rise to a multitude of computer programs generally known in the international literature as multidimensional scaling (MDS). An early review that we still consider one of the most relevant was presented by Michael Subkoviak in 1975 (SUBKOVIAK, 1975).

In Brazil, the technique is known as *escalonamento multidimensional* (multidimensional scaling) (DOS SANTOS; MOREIRA, 1991), but we will use the abbreviation MDS to facilitate reference to the international literature. The first applications to physics education of the version of the technique known today were carried out simultaneously in the second half of the 1970s, at the Institute of Education of Washington University (Missouri, USA) (THRO, 1978) and at the Institute of Physics of UFRGS* (DOS SANTOS, 1978; DOS SANTOS; MOREIRA, 1979; DOS SANTOS; MOREIRA, 1980). Earlier published studies had used word-association tests but not MDS (JOHNSON, 1964, 1967; SHAVELSON, 1973, 1974a). It is important to note that this research area emerged and grew precisely during the expansion of cognitive learning theories. As we shall see below, the two developments are strongly related.

---

* Version of an article published in Portuguese: Felix, H. R. T., Barbosa, G. F., da Costa, J., Rebouças, G. D. O. G., dos Santos, C. A. (2024). Efeitos de aprendizagem na associação de conceitos: um estudo com escalonamento multidimensional. *Revista do Professor de Física*, *8*(1), 269-290.

† TANC is the Portuguese acronym for "Test of Numerical Association of Concepts".

The use of MDS requires advanced computational resources (COXON, 1982; YOUNG, 1987), which were incorporated into the Statistical Package for the Social Sciences (SPSS) only in the late 1990s (DUNN-RANKIN et al., 2004). This may explain the substantial growth, after 2000, in the number of documents retrieved by Google Scholar (Figure 1).

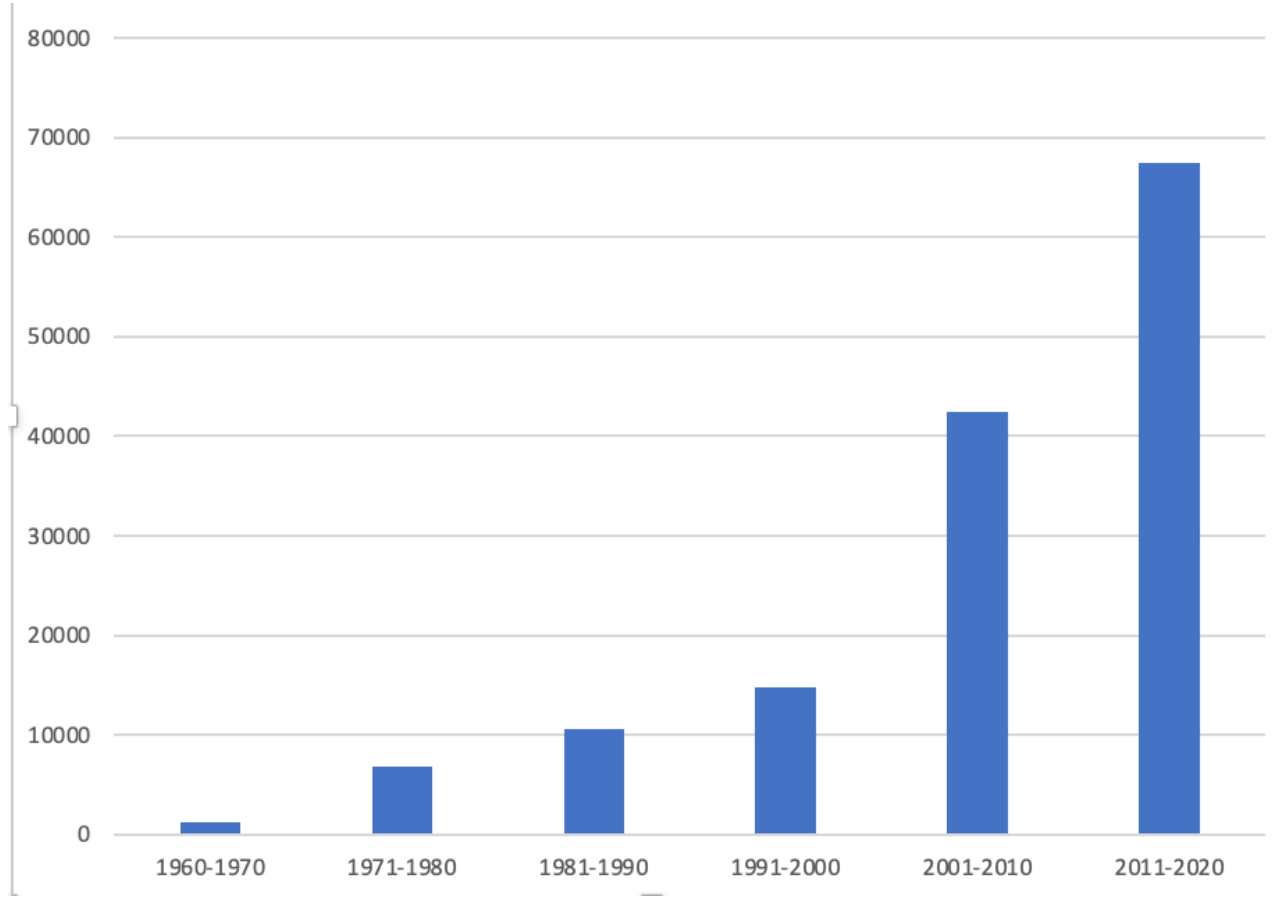

Figure 1: Number of documents retrieved by Google Scholar between 1960 and 2020 for the search phrase "multidimensional scaling".

During an instructional intervention in acoustics for high school students, we observed very interesting results with a concept-association test analyzed using a program that employs the ALSCAL and INDSCAL algorithms (CARROLL; CHANG, 1970; TAKANE; YOUNG; DE LEEUW, 1977; YOUNG; TAKANE; LEWYCKYJ, 1978). Given the quantity and quality of these results, we considered that they should be published separately from the report on the instructional intervention itself.

In the next section, we present a brief review of the literature on the relationship between word-association tests—which we will henceforth call concept-association tests—and current ideas about cognitive structure. We then provide a brief review of contemporary literature on the broad field known as multidimensional scaling, or MDS. The final three sections of the article concern the results obtained in recent months.

* Portuguese acronym for the Federal University of Rio Grande do Sul.

## 2. Concept Association and Cognitive Structure

Following Piaget's studies, the main functional model of cognitive structure was presented by Ausubel (AUSUBEL, 1970; AUSUBEL; FITZGERALD, 1961; AUSUBEL, 1960). Although Bruner (BRUNER, 1960, 1984) made significant contributions to reflections on learning from a cognitive perspective, and is therefore an important reference in this area, he did not develop a model of cognitive structure as functional as Ausubel's. The Brazilian literature on Ausubel's theory is extensive, with Moreira's works (MOREIRA, 1979; MOREIRA, M. A., 2010) among the most important. For this reason, we present here only the basic idea of what Ausubel understands by cognitive structure (AUSUBEL; FITZGERALD, 1961):

> "Cognitive structure" refers to the organization, stability, and clarity of an individual's knowledge in a given field of knowledge (...). In the more general and long-term sense, cognitive-structure variables refer to the influence of significant organizational properties on the acquisition of knowledge in that subject area (...). In the more specific and short-term sense, cognitive-structure variables refer to the effects of the organizational properties of only the immediately (or closely) relevant concepts within a given subject area on the learning and retention of small units of that subject matter.

Underlying this foundation is Ausubel's iconic statement: "Of all the factors influencing learning, the most important is what the learner already knows; ascertain this and teach accordingly" (MOREIRA, 2017). This basic principle of Ausubel's raises a serious question for teachers: what does it mean to ascertain what a student already knows? Many assume that this assessment takes place at the level of knowledge, as measured by ordinary school examinations. Others seek to examine mastery of relevant subsumers (AUSUBEL, 1962) related to the content to be taught, or to identify the existence of alternative conceptions (HAMMER, 1996). Other approaches to accessing cognitive structure are discussed by Tsai and Huang (TSAI; HUANG, 2002). Among these alternatives, we focus on the use of concept-association tests.

Two main concept-association tests are reported in the literature: the Free-Word Association Test (TAEC[*]) (DOS SANTOS; MOREIRA, 1979b; GARSKOF; HOUSTON, 1963) and the Numerical Association of Concepts Test (TANC) (DOS SANTOS; MOREIRA, 1979b; SUBKOVIAK; ROECKS, 1976). The latter, which was used in the present study, simply consists of assigning a value on a Likert scale to the various pairs of concepts. For N concepts, there are $N(N-1)/2$ pairs, assuming that the similarity between concepts A and B is equal to the similarity between B and A. The literature reports the use of Likert scales with different numbers of response alternatives, from 3 to 9. Because it has a midpoint and presents a good level of reliability, we decided to use a scale with 9 alternatives (CHYUNG et al., 2017; PRESTON; COLMAN, 2000).

Shavelson was one of the first researchers explicitly to use concept-association tests to investigate cognitive structure in physics, in the early 1970s (SHAVELSON, 1972, 1974b, a). Underlying this line of research is the question initially raised by Shavelson (SHAVELSON, 1972): "To what extent does the structure of the student's memory, after learning, correspond to the structure of the instructional material?" [Our translation]. Numerous studies have presented results supporting the idea that this correspondence can be examined using concept-association tests (GEESLIN; SHAVELSON, 1975; GLINER, 1981; MOREIRA; DOS SANTOS, 1981; THRO, 1978).

[*] TAEC is the Portuguese acronym for Written Concept Association Test.

From the late 1950s onward, results obtained with word-association tests (in science education, we use the term concept association) began to be analyzed using MDS techniques, initially introduced by Torgerson at the beginning of that decade (TORGERSON, 1952) and refined by several of his successors (KRUSKAL, 1964; SHEPARD, 1962b, a; SUBKOVIAK, 1975). The remarkable results obtained with different types of sensory and psychological stimuli led to the use of tests and MDS in the teaching of physics concepts (DOS SANTOS, 1978). Before reviewing MDS, it may be useful to describe one of the seminal results obtained by Peter Bricker and Sandra Pruzansky at Bell Laboratories (BRICKER; PRUZANSKY, 1970). They produced 24 audible tones (4 frequencies, 3 modulation percentages, and two waveform types, sine and square). The 24 tones were presented to 20 Bell employees. They were then told that they would hear the 552 pairs of tones and should assign a value from 0 to 9 according to the similarity between the tones: "0" for identical tones and "9" for completely different tones. Each tone was heard for 1 second, with a one-second interval between them. After the second tone, respondents had 3 seconds to answer. Bricker and Pruzansky analyzed the results using the MDSCAL program (DOS SANTOS; MOREIRA, 1991), with a three-dimensional Euclidean space. The fitting process identified dimension 1 with frequency, dimension 2 with modulation percentage, and dimension 3 with waveform. Thus, in this and many other cases reported in the literature (DOS SANTOS, 1978) involving psychological and sensory stimuli, word-association tests and MDS computational methods make it possible to identify the physical characteristics of the stimuli.

### 3. Multidimensional Scaling

Given a geographical map, with cities marked by points, it is easy to obtain a table of distances between all the cities shown on the map. One needs only a ruler and the scale used to draw the map. The inverse problem—drawing the map using only the table of distances—does not have such a simple solution. In fact, the solution is quite complex. Multidimensional scaling was created in the 1960s to solve this type of problem.

Following Shepard's seminal work (SHEPARD, 1962a, b), several algorithms were developed in subsequent years, beginning with those devised by Kruskal (KRUSKAL, 1964), Guttman (GUTTMAN, 1968), and Carroll and Chang (CARROLL; CHANG, 1970). Reviews of these algorithms are presented in DOS SANTOS (1978), DOS SANTOS and MOREIRA (1979a, 1991), HOUT, PAPESH, and GOLDINGER (2013), and SUBKOVIAK (1975). In the late 1990s, several of these algorithms were incorporated into the SPSS statistical package. Recent studies conducted in Brazil using SPSS in science education have been reported by GRECA and MOREIRA (2011), HILGER and MOREIRA (2013), and NASCIMENTO et al. (2019).

In the present work, we use a procedure available in SPSS that employs the ALSCAL and INDSCAL algorithms (CARROLL; CHANG, 1970; GIGUÈRE, 2006; TAKANE; YOUNG; DE LEEUW, 1977; YOUNG; TAKANE; LEWYCKYJ, 1978). As indicated in the literature cited so far, the basic objective of the procedure is the same as that of all MDS algorithms: to determine the geometric configuration that best fits a set of proximity measures (similarities or dissimilarities).

There are numerous MDS algorithm models. The classical ones fit data from a single matrix (KRUSKAL, 1964; SHEPARD, 1962a, b; TORGERSON, 1952), but the most widely used are those that fit several matrices and are generally called individual-differences analysis algorithms (CARROLL; CHANG, 1970; TAKANE; YOUNG; DE LEEUW, 1977; YOUNG; TAKANE; LEWYCKYJ, 1978). In addition, algorithms are divided into two other broad categories: metric and nonmetric. As the name suggests, metric algorithms use the values of the proximity measures during fitting. In this respect, they resemble algorithms for fitting experimental data used in the natural sciences and engineering. However, nonmetric algorithms are more widely used; they require only the rank ordering of the proximity measures to determine the configuration that represents them. Some metric algorithms also rank the measures and use strategies from nonmetric algorithms. Among nonmetric algorithms, perhaps the best known is MDSCAL (MultiDimensional SCALing), developed by Kruskal (KRUSKAL, 1964), which inspired almost all algorithms developed in the 1960s and 1970s.

The basic principle of these algorithms is almost always the same. Their differences essentially concern the mathematical and computational strategies used to obtain the configuration ($d_{ij}$) that best fits the similarity measures ($s_{ij}$). This configuration is generally obtained in a Euclidean space that may have any number of dimensions, R, and not merely two:

$$d_{ij} = \sqrt{\sum_{r=1}^{R}\left(x_{ir} - x_{jr}\right)^2} \quad (1)$$

The fitting criterion is defined by the stress function, S, as defined by Kruskal:

$$S = \sqrt{\frac{\sum_{i,j}\left(d_{ij}-\hat{d}_{ij}\right)^2}{\sum_{i<j} d_{ij}^2}} \quad (2)$$

where $\hat{d}_{ij}$ are numbers monotonically related to $s_{ij}$. That is, if $s_{ij} < s_{kl}$, then $\hat{d}_{ij} < \hat{d}_{kl}$. To obtain the best configuration, the algorithm must minimize S. This process is similar to the least-squares method, which is well known to those who work with fitting experimental data.

Let us try to understand this basic principle through an intuitively simple result. Although, in real cases, fitting can be performed in several dimensions, for simplicity we will consider a two-dimensional fit. We constructed a matrix with the straight-line distances between 20 Brazilian state capitals, available at https://goodway.com.br/distancias.htm, and fitted it using ALSCAL in SPSS. As expected, we obtained a “map of Brazil” (Figure 2).

Superimposing the obtained points on the actual map so that they “coincide” with the locations of the state capitals is not always immediate—in fact, it is almost never immediate. Two types of geometric operations are required for the “obtained map” to coincide with the “real map”: expanding (or shrinking) and rotating the obtained map until its points lie on, or very close to, the points on the other map. These operations do not change the rank ordering of proximities.

Therefore, MDS solutions are invariant under these operations. All that needs to be preserved is the rank ordering of proximities.

This example illustrates the algorithm's ability to correctly represent a geometric configuration of N points from the N(N−1)/2 distances between them.

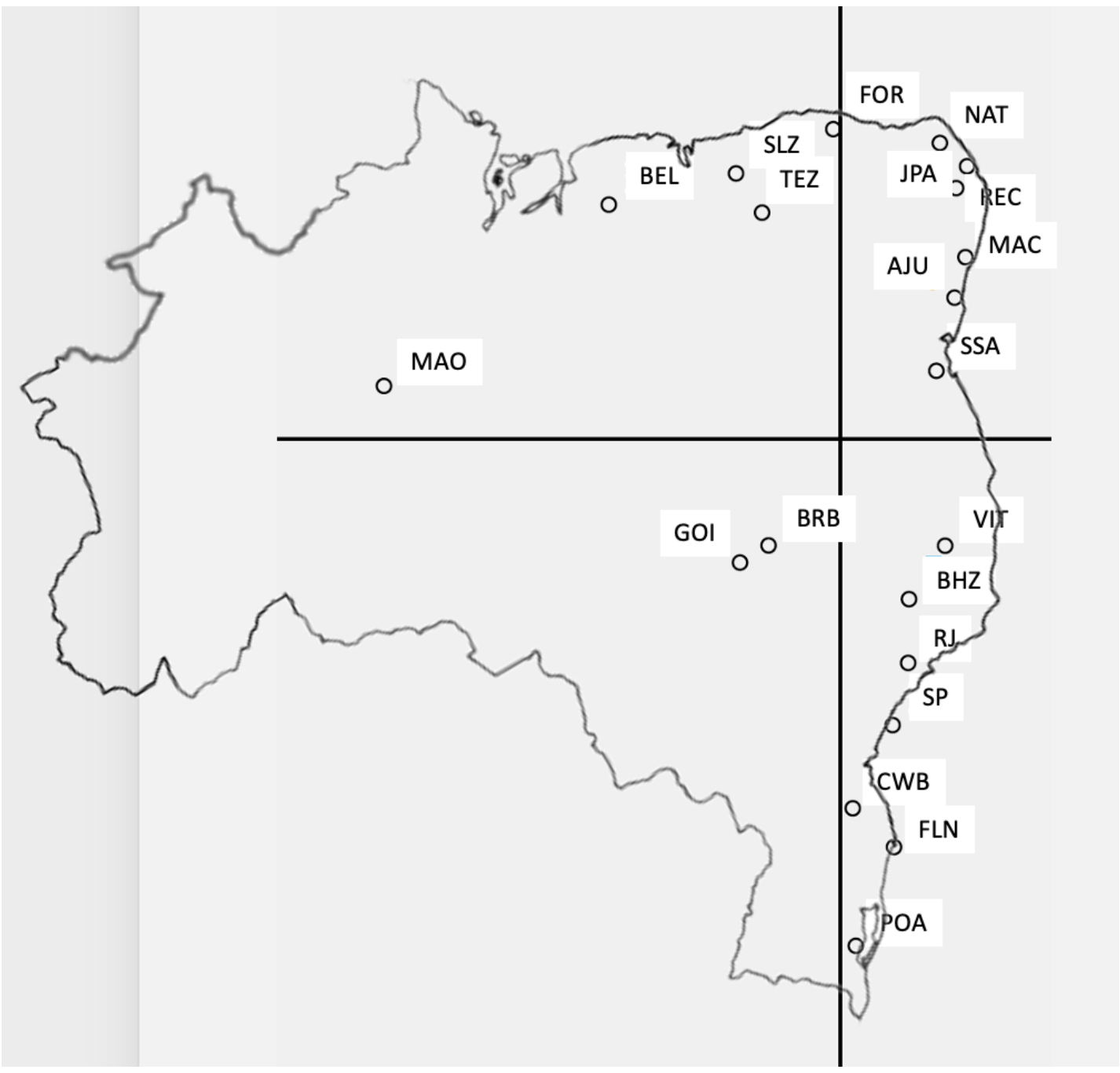

Figure 2: Part of the map of Brazil obtained with ALSCAL/INDSCAL from aerial distances between 20 state capitals.

However, in the case of sensory, psychological, or cognitive stimuli, one may still wonder whether the geometric representation actually corresponds to the similarity measures. In other words, how does this representation differ from one generated randomly? A conceptual map obtained from a random matrix is shown in Figure 3. We will show later that, unlike this random map, all maps obtained with the TANC present, in one way or another, a configuration consistent with the structure of the content addressed in the classroom.

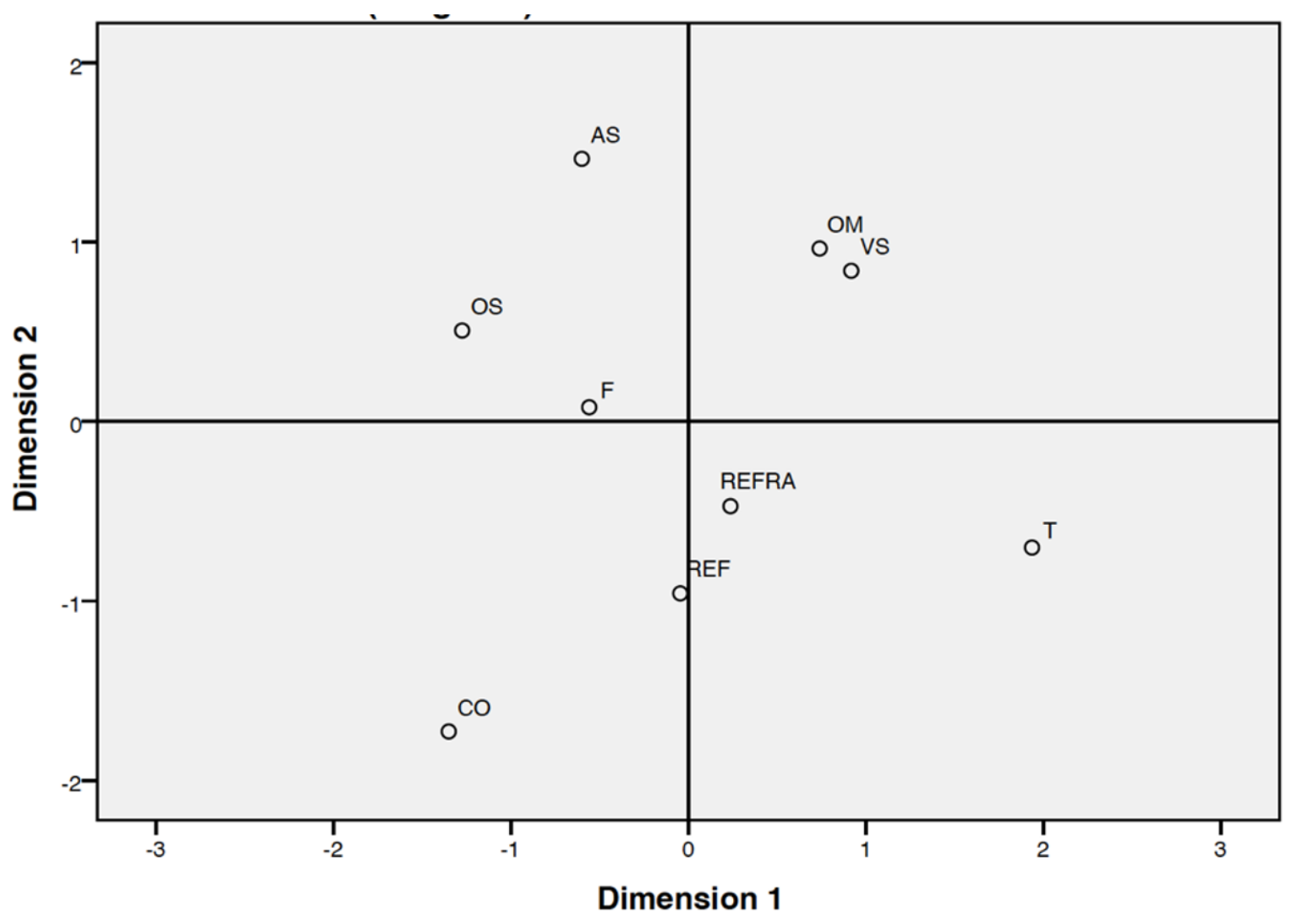

Figure 3: Map obtained from a random matrix.

How does the algorithm arrive at the solutions shown in Figures 2 and 3? The procedure involves four steps. Assuming that the proximity matrix has N(N−1)/2 cells, corresponding to the proximities of N objects, the algorithm proceeds as follows (SUBKOVIAK, 1975):

1. It starts with an arbitrary configuration.

2. It calculates the distances between all points in the configuration (Equation 1) and ranks them in ascending order.

3. It calculates the stress function (Equation 2).

4. It returns to Step 1 and rearranges the configuration to reduce the value of S. To achieve monotonicity and improve the fit, points that are farther apart than the similarity measure indicates must be brought closer together, while points that are closer than the similarity measure indicates must be moved farther apart.

Steps 1–4 are repeated several times until the decrease in S becomes negligible. In ALSCAL/INDSCAL, the process is stopped when the decrease in S is on the order of 0.00001. Figure 4 illustrates the process of changing distances to satisfy monotonicity. Figure 4a shows a perfect monotonic relationship between distances and similarities, which is almost impossible to obtain in real cases. In the stage of the fitting process represented in Figure 4b, in order to reduce S, points (3,4) and (1,2) must be brought closer together, while points (1,3) and (1,4) must be moved farther apart.

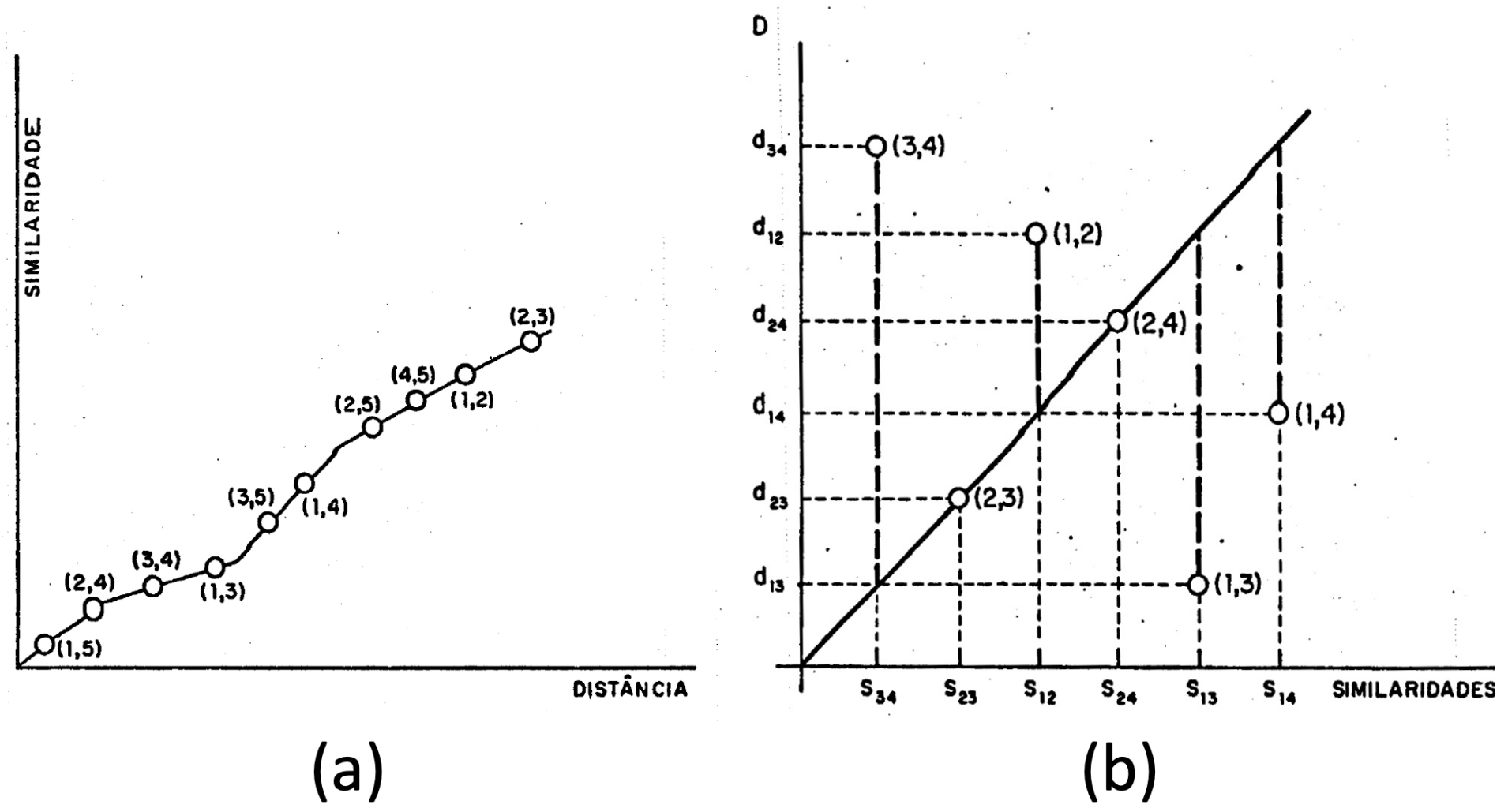

Figure 4: (a) Perfect monotonic relationship between distances and similarities; (b) incomplete monotonic relationship. Images reproduced from DOS SANTOS (1978).

In individual-differences algorithms such as INDSCAL, a Euclidean space is obtained for each subject who provides proximity measures. Since each individual has their own criteria for evaluating proximities, a weighted Euclidean space is obtained. That is, for each individual k, a point configuration is obtained whose Euclidean distances between points i and j, $d_{ijk}$, are given by:

$$d_{ij}^{k} = \sqrt{\sum_{r=1}^{R} w_{kr}\left(x_{ir} - x_{jr}\right)^{2}}, \tag{3}$$

where $w_{rk}$ represents the weight, or importance, that subject k assigns to dimension (or factor) r. In this case, the function fitted by the program is similar to the stress defined by Kruskal and, for a sample of m subjects, has the following form:

$$f(X_1, \dots X_m) = \frac{1}{m}\sum_{k=1}^{m}\sum_{i<j}^{n} w_{ij}^{k}\left[\delta_{ij}^{k} - d_{ij}^{k}\right]^{2}. \tag{4}$$

The INDSCAL procedure in SPSS provides a stimulus matrix corresponding to a kind of average of the group of respondents, as well as a plot of the respondents' weights. To obtain the individual map for each respondent, the matrices must be run individually, as we will see below.

## 4. Methodology of the Present Study

The study was conducted students enrolled in a third-year high school class at a school in the state of Ceará with a low socioeconomic status index (INSE*). The class consisted of 22 male and 11 female students, aged between 17 and 19. In November 2022, they attended five conventional classes on basic wave phenomena and the physiology of sound, using the school textbook (PIETROCOLA et al., 2016). In June 2023, they completed a TANC with nine concepts—sound wave (OS), mechanical wave (OM), timbre (T), frequency (F), wavelength

* INSE is the Portuguese acronym for socioeconomic status index.

(CO), reflection (REF), sound amplitude (AS), speed of sound (VS), and refraction (REFRA)*—and a nine-point Likert scale. From this point onward, this test will be called the “pre-test.” Table 1 presents the instructions and part of the TANC.

Table 1. Numerical Association of Concepts Test (TANC).

This test is designed to assess the degree to which you perceive a relationship or association between certain concepts in wave physics and acoustics. The following page contains a number of pairs of concepts. Next to each pair is a numerical scale ranging from 1 to 9. On this scale, 1 represents a high degree of relatedness between the concepts (for example, temperature and heat, or mass and energy), while 9 represents virtually no relationship (for example, volume and velocity). If you believe that the concepts in a given pair are strongly related, place an X in column 1 of the scale. Conversely, if you believe that the two concepts have nothing to do with each other, place an X in column 9. These represent the two extremes. The numbers between 1 and 9 reflect intermediate positions: the closer your X is to 1, the stronger the relationship you perceive between the concepts; the closer it is to 9, the weaker that relationship.

(There is no time limit for this test)

| | 1 | 2 | 3 | 4 | 5 | 6 | 7 | 8 | 9 |
|---|---|---|---|---|---|---|---|---|---|
| Sound wave - Mechanical wave | | | | | | | | | |
| Timbre – Onda mecânica | | | | | | | | | |
| Timbre – Mechanical wave | | | | | | | | | |
| Mechanical wave - Frequency | | | | | | | | | |
| Wavelength - Frequency | | | | | | | | | |
| Sound wave - Wavelength | | | | | | | | | |
| Timbre – Reflection | | | | | | | | | |
| Frequency - Sound amplitude | | | | | | | | | |
| Wavelength - Mechanical wave | | | | | | | | | |
| - - - - - | | | | | | | | | |

Between August and September 2023, the students attended four classes using Ausubelian texts on acoustics and music. The instructional procedures in this intervention, henceforth called the Theoretical Instructional Intervention (IDT), will not be discussed in this paper. At the end of the IDT, they completed the TANC again, referred to as the “post-IDT test.” At the end of September, they attended four classes on how the Arduino works, assembled a piano keyboard, and programmed some songs. During this stage, they revisited acoustics concepts in the context of musical practice. This stage will be called the Experimental Instructional Intervention (IDE). At the end of the IDE, they completed the TANC for the third time, here called the “post-IDE test.”

The IDT and IDE instructional interventions will be presented on another occasion. The objective of the present study is to investigate whether conceptual appropriation during a given

* Onda sonora (OS), onda mecânica (OM), timbre (T), frequência (F), comprimento de onda (CO), reflexão (REF), amplitude do som (AS), velocidade do som (VS), and refração (REFRA).

instructional intervention can be identified through the TANC when analyzed using ALSCAL/INDSCAL. The class underwent three different instructional stages, with the TANC administered after each one. The question to be examined is: do the TANC results depend on these interventions?

As a reference, we use the results obtained by administering the TANC to four physics teachers. Three of these teachers hold PhDs in condensed-matter physics, but acoustics is not their research area and they have no high school teaching experience. The third respondent is a high school teacher and a master's student in the National Professional Master's Program in Physics Teaching (MNPEF-UFERSA). He was responsible for implementing the instructional interventions.

MDS fits are generally performed in a two-dimensional space, although three-dimensional fits are sometimes presented, and fits with more than three dimensions are rarely reported in the literature. Some criteria for determining the most appropriate dimensionality are presented in DOS SANTOS (1978). Depending on the data, a simple examination of the resulting maps makes it possible to choose the dimensionality. In the results obtained in the present study, we found that three-dimensional fits provide no information beyond that obtained from two-dimensional fits. In addition, for a three-dimensional analysis, the minimum number of points should be 13. Empirical results have shown that $N = 4r + 1$ (SUBKOVIAK, 1975). Therefore, for $N = 9$, $r = 2$.

Another important question concerns the meaning of the dimensions. In the example mentioned above, involving the results obtained by Bricker and Pruzansky (BRICKER; PRUZANSKY, 1970), the stimuli (tones) used in that experiment had well-defined physical parameters: frequency, modulation percentage, and waveform type. The MDSCAL fit clearly showed that the three dimensions were associated with these physical parameters. In the case of psychological and scientific concepts, their complex relationships make it difficult, if not impossible, to identify the meaning of the dimensions. In other words, in many of these more complex cases, dimensions are no more than computational artifacts used to obtain the configuration. For this reason, we decided to use a topological metaphor to interpret the results obtained in this study. Instead of dimension 1 versus dimension 2, we use the cardinal directions: North, South, East, and West.

## 5. Results

### *5.1 Teachers' maps*

The teachers' maps are presented in Figure 5. Figure 5b indicates that Teacher 3 uses criteria for associating concepts that differ from those used by the other teachers. This result seems to be related to the fact that Teacher 3 is a high school teacher and was responsible for implementing the instructional interventions, whereas the other three teachers hold PhDs in physics but have no teaching experience in waves and acoustics at the high school level. On the other hand, throughout their academic careers, from undergraduate studies through the master's and doctoral degrees, each of these teachers has encountered wave concepts from different perspectives. The electromagnetic spectrum, notions of atomic physics, and the study of different approaches to wave theory lead these teachers to regard frequency and wavelength as interchangeable. Thus, while many students encountering wave physics for the first time associate speed, frequency, and wavelength through the fundamental relationship

$$v = \lambda f, \tag{5}$$

those who have had the opportunity for more advanced study assimilate the idea that frequency and wavelength can be interchangeable. This may be why these three concepts form a triangle in Teacher 3's map (Figure 5c) and overlap in Teacher 4's map (Figure 5d). This notion of interchangeability does not appear in the students' case, as we will see below. The topologies of some students' maps are similar to that presented in Teacher 3's map.

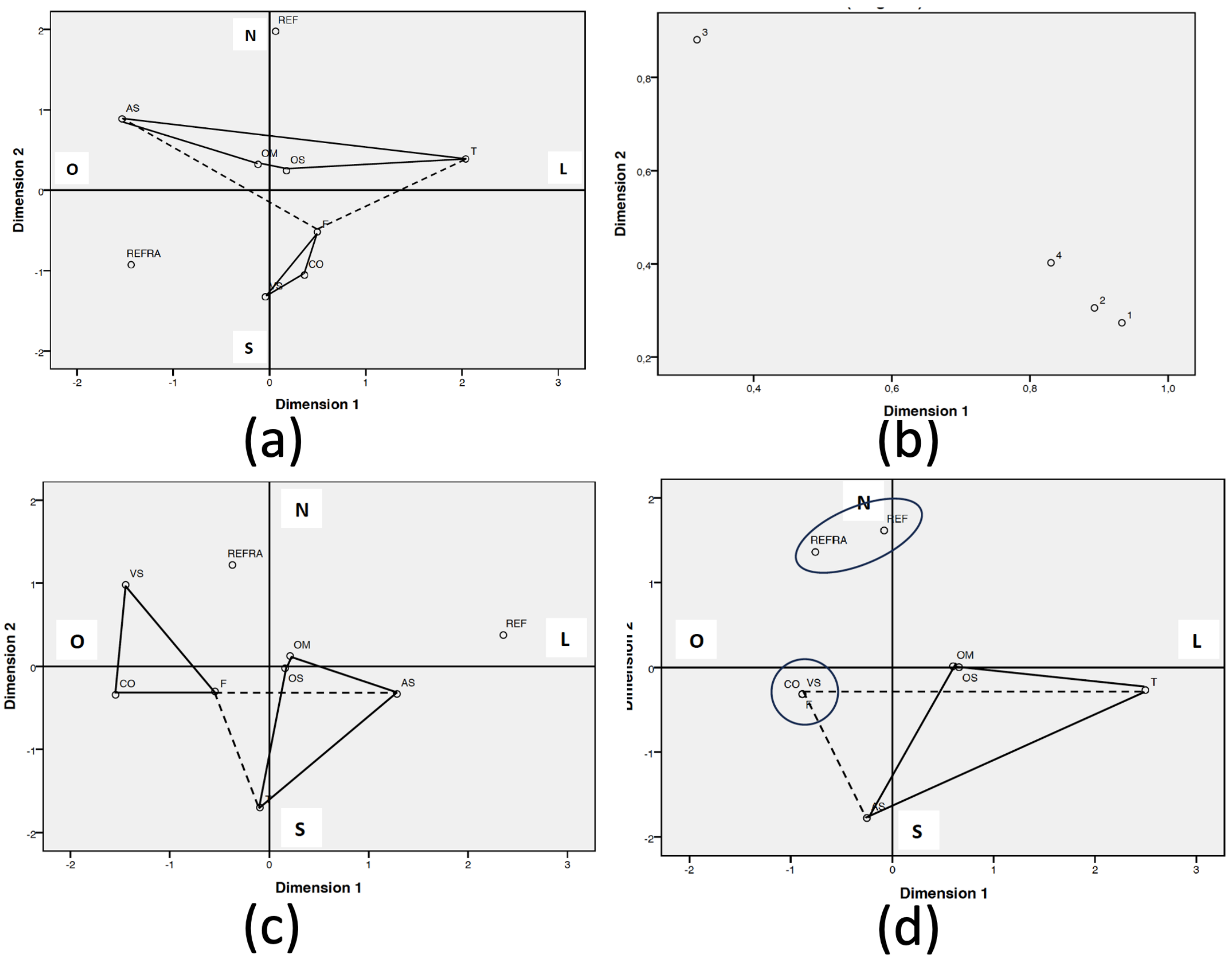

Figure 5: (a) Teachers' map; (b) weights plot; (c) Teacher 3's map; (d) Teacher 4's map. The labels N, L, S, and O represent cardinal directions in Portuguese: Norte (N) [North], Leste (L) [East], Sul (S) [South], and Oeste (O) [West].

The topology of the overall map (Figure 5a) makes it possible to distinguish three groups of concepts. The first follows from Equation 5: OO, VS)*. The second concerns acoustics and the

* This order is not arbitrary. It follows the clockwise direction in the geometric figure formed by the concepts. This is relevant from the topological point of view and with respect to invariance under rotations. Note that the order of the concepts in the group (T, OS, OM, AS) follows this topological criterion.

physiology of sound: (T, OS, OM, AS). The third group (F, AS, T) concerns the definition of timbre, the characteristic of sound that enables us to distinguish two signals with the same frequency and amplitude but produced by different sound sources. In many maps, REFRA is close to VS, which may be related to the law of refraction:

$$n_1 sen\theta_1 = n_2 sen\theta_2 \,, \tag{6}$$

where $n_1 = \frac{c}{v_1}$ e $n_2 = \frac{c}{v_2}$ are the refractive indices of the two media.

Finally, the relative positions of REF and REFRA deserve a more detailed discussion. The fact that the phenomena of reflection and refraction do not explicitly share any of the concepts presented in the TANC may explain the large separation observed in some teachers' and students' maps. On the other hand, cases in which these concepts are close (Figure 5d) may reflect two things: simply a semantic association, or the perception that they are phenomena observed in all types of waves.

*5.2 Students' maps: pre-test*

The students' map obtained from the pre-test is shown in Figure 6a, while the dimensional weights are shown in Figure 6b. Comparing Figures 5a and 6a, we see that students use criteria for concept association that differ from those used by the teachers. The group (REF, REFRA) may be due to a semantic association, as may the group (OM, CO, OS). The group (F, T, AS), in turn, may be a remnant of the classes on the physiology of sound that the students had attended seven months before the TANC was administered. It should be noted that the concept VS is clearly isolated from the other concepts, especially F and CO.

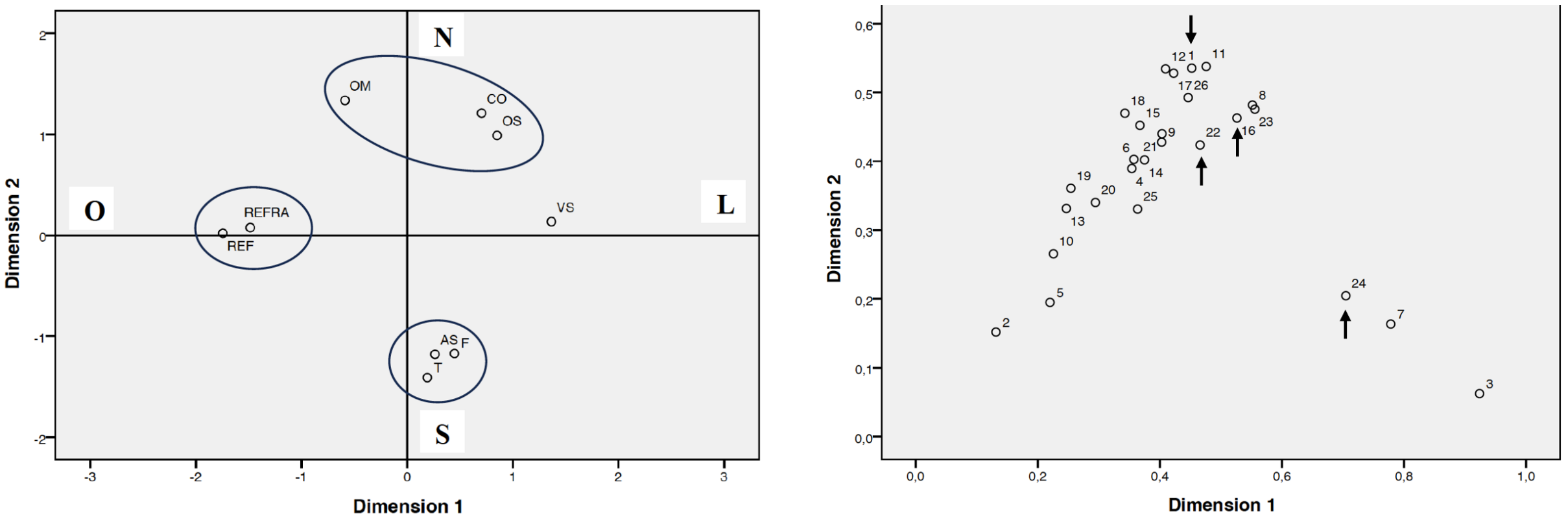

(a) (b)

Figure 6: (a) Students' map obtained from the pre-test; (b) distribution of dimensional weights.

In Figure 6b, four students are highlighted. They were selected by the teacher responsible for implementing the instructional intervention on the basis of their school performance (ordinary examinations) and their participation in classroom activities throughout the implementation of the educational product. Figure 6b suggests that Students 1, 16, and 22 use similar criteria to define the associations between the concepts presented in the TANC, whereas Student 24, whose weight on dimension 1 is three times that assigned to dimension 2, appears to use different criteria. In analyzing the pre-test results, it would be interesting to examine how much differences in weights influence topological structures, as shown in Figure 7. At first glance, these are four topologically different maps. None can be obtained from any of the others through rotations. There are traces of learning from when the students studied the subject six months earlier, but each map presents these traces differently. What, then, is the effect of the weights on the topology of the maps?

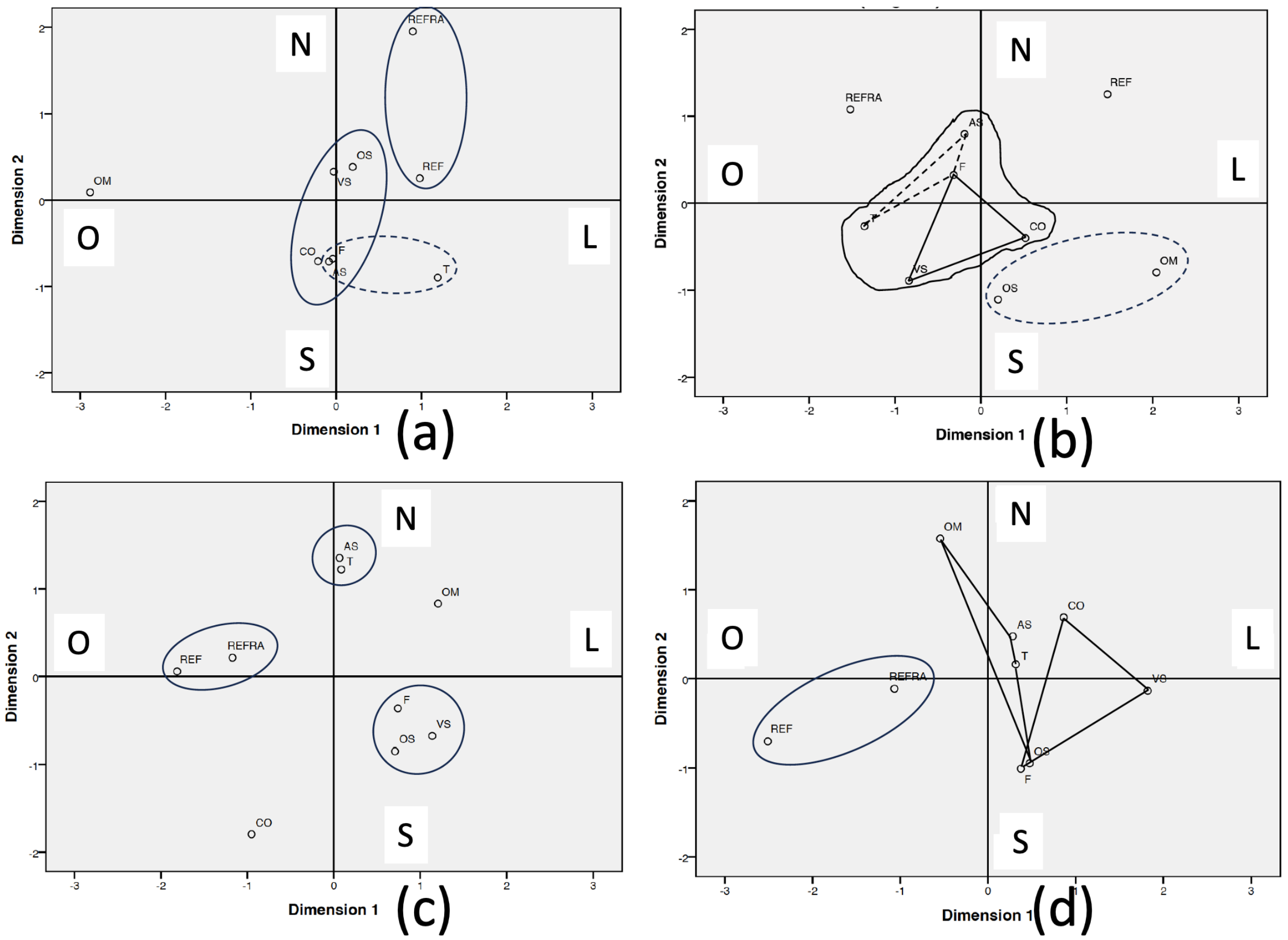

Figure 7: (a) Student 1's pre-test map; (b) Student 16's pre-test map; (c) Student 22's pre-test map; (d) Student 24's pre-test map.

Let us begin with Student 22's map, the most different of all. Everything indicates that semantic criteria predominated in the association of concepts. Equations 5 and 6 do not appear to be part of the criteria used by this student. The strong similarity between timbre and sound amplitude probably comes from common sense. If it came from the definition of timbre, the concept of frequency should be closer to the pair (T, AS). The formation of the group (F, VS, OS) must also have been guided by semantic criteria. Therefore, to use a simple definition, we can say that this is a predominantly semantic map.

Student 1 (Figure 7a) also presents a strong semantic component, although scientific criteria guided the formation of some groups. For example, the concepts related to sound are all grouped in the center of the map, while the concept of mechanical wave seems not to be part of the area of study. The concepts of frequency and sound amplitude are closer to wavelength than to timbre, suggesting that semantic criteria take priority in concept association. Nevertheless, Equation 5 is present in the vicinity of F, VS, and CO. Thus, Student 1's map can be classified as a semantic map with elements of a scientific map.

Students 16 and 24 are among those with the best academic performance, and they were also the students who participated most in classroom activities. Their mental structures, prior to the instructional interventions, are similar to that presented by the teacher responsible for the course (Figure 5c). Both maps present the same groups: (F, CO, VS), (F, AS, T), and (T, OS, OM, AS). A rotation L–N–O provides congruence between the student's and the teacher's groups (F, CO, VS). The students' groups (F, AS, T) and (T, OS, OM, AS) have topologies similar to the teacher's through an S–O rotation. A notable aspect of these similarities is that the groups (F, AS, T) and (F, CO, VS) are contiguous in both maps and form a large group that can be associated with the set of wave properties. Therefore, these two maps appear to have topologies formed from similarity measures between concepts assigned predominantly according to scientific criteria.

Examining Figure 6b leads us to suppose that the weight on dimension 2 is related to semantic criteria, whereas the weight on dimension 1 is related to scientific criteria. The closer a point in Figure 6b is to high values on dimension 1, the more important the scientific criteria used by the respondent. However, the algorithm's resolution does not allow us to place complete confidence in this hypothesis. For this reason, we consider it more reliable to interpret the maps on the basis of the topology of the concept groups.

### *5.3 Students' maps: post-tests*

The students' maps obtained after the IDT (post-test 1, PT1) and the IDE (post-test 2, PT2) are presented in Figures 8a and 8b, respectively. There are notable differences between the maps in Figures 6a (pre-test), 8a (PT1), and 8b (PT2). The semantic effect attributed to the pre-test configuration, if it did not disappear, must have decreased considerably in PT1. Apparently, the perception that OM and OS are entities of the same nature became consolidated. In addition, the fundamental relationship (Equation 5) is clearly displayed in the group (F, CO, VS). Likewise, the acoustic properties are highlighted by the group (F, AS, T), which appears in Figure 8a with the topology (F, T, AS). The two topologies can be brought into congruence by a projection along the N–S axis.

Figure 8b suggests that the IDE produced a reorganization of the students' average mental map, consistent with the instructional emphasis placed on acoustics concepts. In the IDE, students assembled an electronic keyboard connected to an Arduino board and programmed the board to play some songs. Therefore, throughout the IDE they worked in considerable detail with the concepts of timbre (T), frequency (F), and sound amplitude (AS). They also discussed extensively the differences and similarities between mechanical waves in general and sound waves.

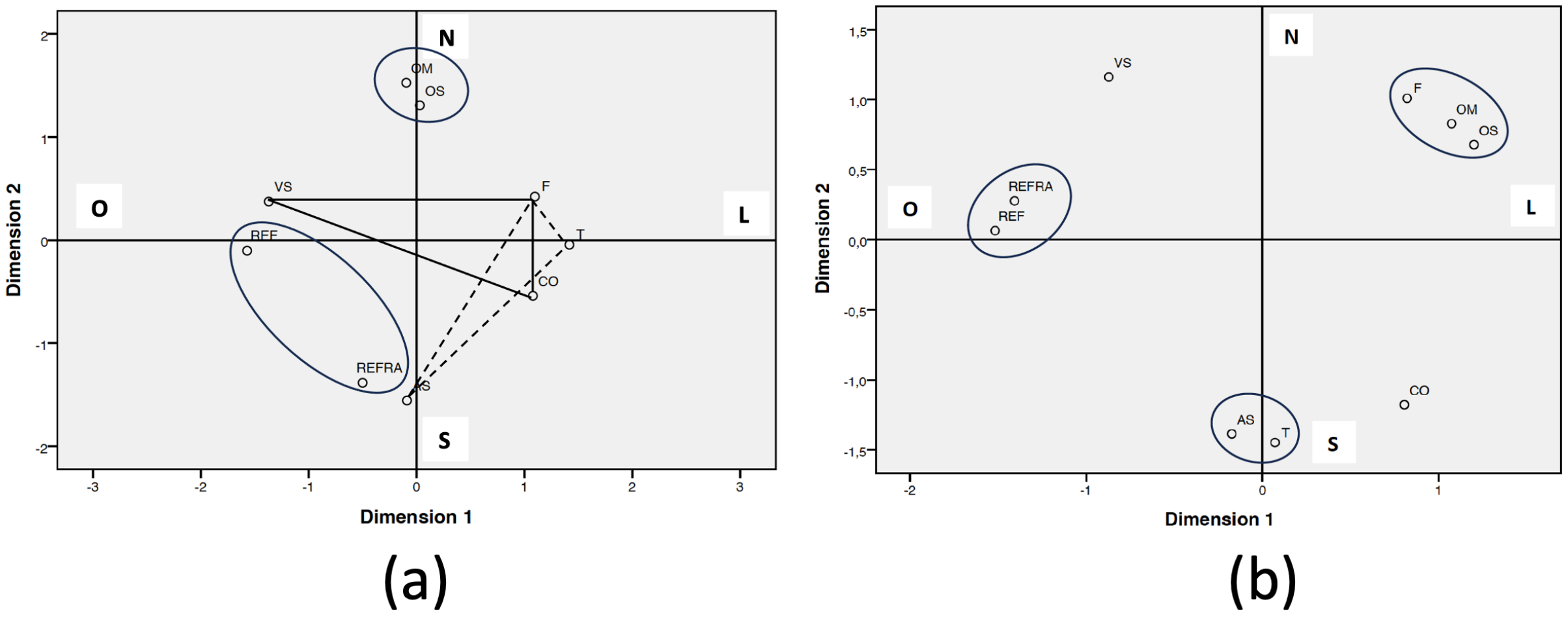

Figure 8: (a) Students' map obtained from post-test 1; (b) students' map obtained from post-test 2.

The maps of the same students mentioned above, obtained in PT2, are presented in Figure 9. Comparing Figures 7 and 9, we see that all four students changed their mental configurations after the instructional intervention. In the pre-test, Student 1 (Figure 7a) associated the concepts of reflection and refraction in the way that is customary, even in common-sense thinking. When someone studies acoustic phenomena, reflection is a more important concept than refraction. This perception may be one reason for the separation of the group (REF, REFRA) in the map in Figure 9a. In the study of acoustics, the sound wave is a fundamental element, and in a conceptual map this element attracts the properties pertinent to it. Thus, from an epistemological point of view, the formation of the group (OM, OS, AS) makes sense. It should be noted that OM, which was displaced in Figure 7a, was drawn into this group, which we may call the acoustic group. The conceptual understanding of timbre is strongly related to the frequency and amplitude of a sound wave, but has a weak direct relationship with the speed of sound. This may have moved the concept away from its natural group (T, F, AS, OS) in Figure 9a.

The results suggest that Student 16 consistently uses scientific criteria in concept association, but the comparison between Figures 7b and 9b suggests a strong influence of the IDE on these criteria. The persistence of the acoustic group (T, AS, F) and the splitting of (F, CO, VS) in PT2 suggest that Equation 5 lost importance in comparison with the foundations of acoustics. This behavior is also suggested by the movement of OS toward the acoustic group.

Student 22 showed a significant change in his mental map after the IDE (Figure 9c). The isolation of the concept of refraction and the movement of reflection toward concepts typically associated with acoustics are strong indications of this change, guided by scientific criteria. Another strong indication is the consolidation of the groups (T, AS, F) and (F, VS, CO).

The most unexpectedF result was presented by Student 24. In the pre-test (Figure 7d), his map contained the groups (F, CO, VS) and (T, OS, OM, AS), which were split apart in PT2. The most curious aspect is that, in the pre-test, the typically musical group (T, F, AS) was located between (F, CO, VS) and (T, OS, OM, AS), whereas in PT2 the concept AS is separated from (T, F).

Because of this apparently anomalous result, we decided to interview this student. The most important information from the interview, conducted by the course teacher, is presented below.

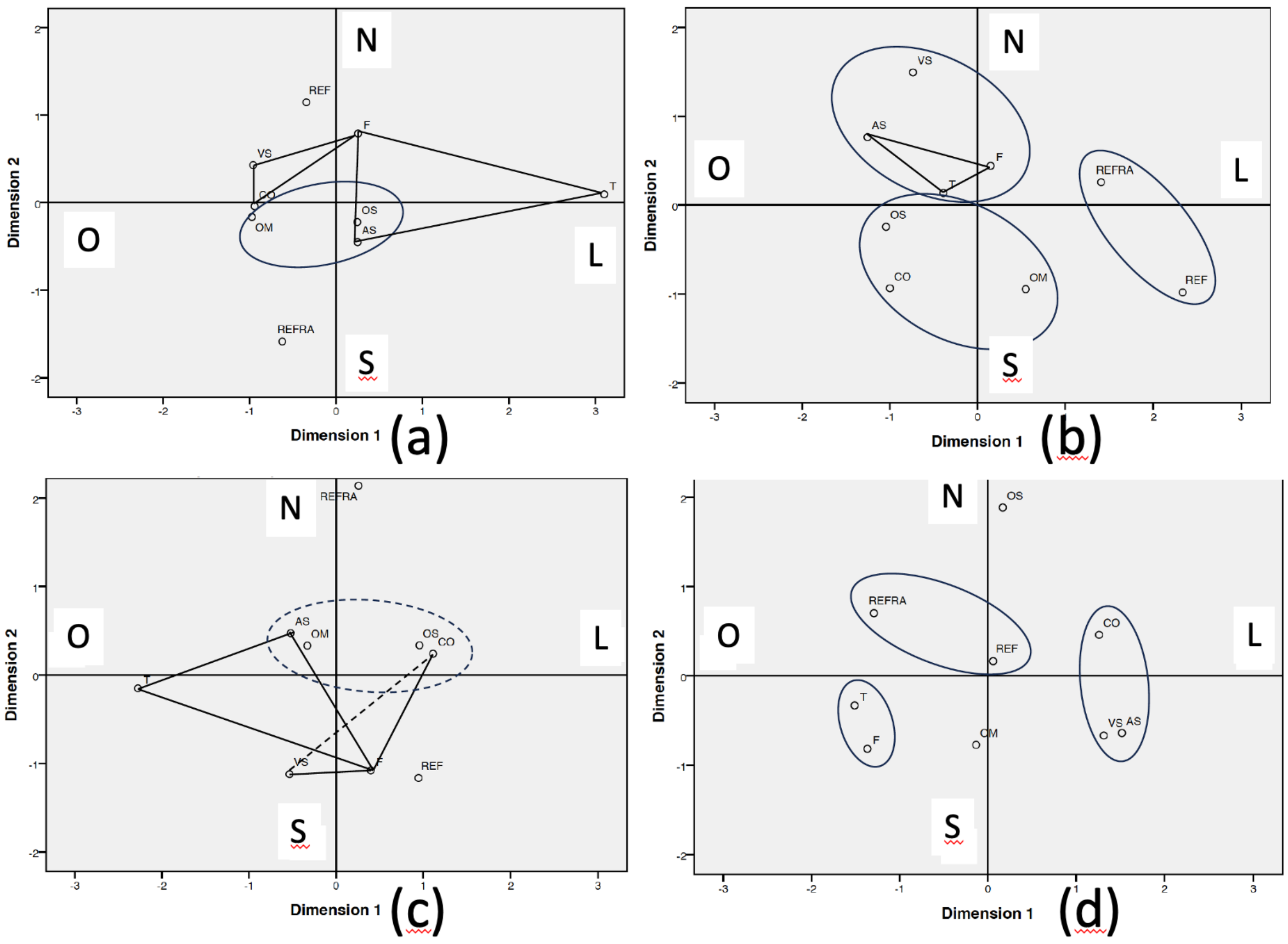

Figure 9: (a) Student 1's PT2 map; (b) Student 16's PT2 map; (c) Student 22's PT2 map; (d) Student 24's PT2 map.

The maps in Figure 9 were shown to the student without the markings of the concept groups. After presenting the maps, the teacher asked the following questions: (a) Which of the maps below did you produce in post-test 2? (b) Can you explain the map?

After examining the maps in detail, the student identified Student 16's map (Figure 9b) as his own. This was more or less what we expected. In view of Student 24's progress in the course, we expected his map to be similar to Student 16's. In explaining his choice, he stated that F is closer to T. He also highlighted the proximity of REFRA and F.

When the teacher informed him that he was Student 24 (Figure 9d), he realized that his map was incorrect and admitted that he had not been very careful in answering PT2. When commenting on his mistakes, he displayed knowledge more consistent with Student 16's map. For example, he stated that F and CO should be very close "because they are inversely related." During the final course activities, students played songs with the aid of an Arduino board. This was probably why Student 24 stated that T and F should be closer than they were in his map, as should CO. Once again, Student 24 showed that he had internalized Equation 5 when he stated that F and CO should be closer because "one affects the other; the higher the frequency, the shorter the wavelength, if I am not mistaken."

Examination of Figure 9d and Student 24's statement suggests two conclusions. When analyzed through multidimensional scaling, TANC is capable of projecting an individual's cognitive structure. This is demonstrated by the fact that we can assess the accuracy of the student's conceptual map on the basis of what we know about his performance in the classroom. The other conclusion is that concept association is strongly influenced by learning.

*6. Final comments*

An unconventional instructional intervention in the area of wave phenomena was applied to a group of high school students six months after the class had studied the same content through the usual instructional approach, namely conventional lessons on the topic covered in the textbook and knowledge tests of the type used in ENEM[*]. The instructional intervention was presented in two stages. In the first, called the Theoretical Instructional Intervention (IDT), part of the content presented in the earlier approach was revisited using an instructional structure inspired by Ausubel (AUSUBEL, 1970; AUSUBEL, 1960). Two advance organizers ("Acoustics in Our Daily Lives" and "Music and Physics") were read and discussed by the students in the classroom during the IDT. In the second stage, called the Experimental Instructional Intervention (IDE), students worked with an Arduino board, built a piano keyboard, and played some songs they had programmed. At the end of the IDE, the teacher conducted a lesson on integrative reconciliation, in accordance with Ausubel's learning theory (AUSUBEL, 2000; MOREIRA, 1979; MOREIRA, 2010).

In addition to the usual knowledge tests, students answered two types of tests. Before the IDT and after the IDE, they took a knowledge test to examine the existence of alternative conceptions. Before the IDT, students answered a Numerical Association of Concepts Test (TANC) (DOS SANTOS; MOREIRA, 1980), referred to here as the pre-test. The same test was administered after the IDT (post-test 1, PT1) and after the IDE (PT2). The concept-association tests, which are the subject of the present work, were analyzed using ALSCAL, a multidimensional scaling program (DOS SANTOS; MOREIRA, 1991) available in the SPSS statistical package.

The results presented here are consistent with reports in the literature concerning concepts in mechanics (SHAVELSON, 1972), thermodynamics (DOS SANTOS; MOREIRA, 1979b; MOREIRA; DOS SANTOS, 1981), and quantum mechanics (GRECA; MOREIRA, 2011). All these reports show strong indications that concept-association tests, when analyzed using multidimensional scaling algorithms, reflect the respondent's cognitive structure, and that this structure reflects the structure of the subject matter studied. These algorithms make it possible to display the typical cognitive structure of a group of respondents—something like an average cognitive structure—as well as individual cognitive structures. In the process of fitting test results to a representative geometric configuration, the algorithms determine the weights that each respondent assigns to each dimension. These weights are related to the importance of the dimension in the criteria used to evaluate similarity measures. In the present study, after a detailed analysis of the topologies of the groups shown in the maps and of plots of the weights, such as Figure 6b, we formulated the hypothesis that dimension 2 may be related to semantic criteria, whereas dimension 1 may be related to scientific criteria. However, the algorithm's resolution does not allow us to place complete confidence in this hypothesis. For this reason, we consider it more reliable to interpret the maps on the basis of the topology of the concept groups.

[*] Portuguese acronym for the Brazilianufrgs National High School Exam.

The changes observed in the maps before and after the two stages of the instructional intervention are consistent with the structure of the content addressed. The only anomaly was presented in Student 24's PT2 (Figure 9d), but the interview we conducted with the student convinced us that it was an apparent anomaly. The student admitted that he had been careless when answering PT2.

In conclusion, it is important to bear in mind that the result reported here refers to the specific case of our instructional intervention, which consisted of two stages (IDT and IDE). It would not be prudent to generalize it, but it would be interesting for similar trials to be conducted with different subject matter and different instructional interventions.